\documentstyle[epsf,aps]{revtex}

\def\beq{\begin{equation}}
\def\eeq{\end{equation}}
\def\bea{\begin{eqnarray}}
\def\eea{\end{eqnarray}}
\def\non{\nonumber}

\newcommand{\eq}[1]{eq.~(\ref{#1})}

\def\np#1#2#3{Nucl.\ Phys.\ {\bf B#1}, #2 (19#3)}
\def\pl#1#2#3{Phys.\ Lett.\ {\bf B#1}, #2 (19#3)}
\def\pr#1#2#3{Phys.\ Rev.\ {\bf D#1}, #2 (19#3)}
\def\prl#1#2#3{Phys.\ Rev.\ Lett.\ {\bf #1}, #2 (19#3)}

\def\zp#1#2#3{Zeit.\ Phys.\ {\bf C#1}, #2 (19#3)}

\def\mafigura#1#2#3#4{
  \begin{figure}[htbp]
    \begin{center}
      \epsfxsize=#1
      \leavevmode
      \epsffile{#2}
    \end{center}
    \caption{#3}
    \label{#4}
  \end{figure} }

\begin{document}
\preprint{FTUV/97-16}
\preprint{IFIC/97-16}
\preprint{PRA-HEP 97/5}

\title{Dimensionally Regularized Box and Phase-Space 
Integrals Involving Gluons and Massive Quarks}

\draft 

\author{Germ\'an Rodrigo and Arcadi Santamaria}
\address{Departament de F\'{\i}sica Te\`orica, IFIC,
CSIC-Universitat de Val\`encia, 46100 Burjassot, Val\`encia, Spain} 
\author{Mikhail Bilenky\thanks{On leave from JINR, 141980 Dubna, 
Russian Federation}}
\address{Institute of Physics, AS CR, 18040 Prague 8, and 
Nuclear Physics Institute, AS CR, 25068 \v{R}e\v{z}(Prague), Czech Republic}

\date{March 18, 1997}

\maketitle

\begin{abstract}
The basic box and phase-space integrals needed to compute
at second order the three-jet decay rate of the $Z$-boson into
massive quarks are presented in this paper.
Dimensional regularization is used to regularize the 
infrared divergences that appear in intermediate steps.
Finally, the cancellation of these divergences 
among the virtual and the real contributions
is showed explicitly.
\end{abstract}

\pacs{12.15.Ff, 12.38.Bx, 12.38.Qk, 13.38.Dg, 14.65.Fy}

\section{Introduction}

In higher order QCD calculations at LEP
the effects of quark masses 
are always screened by the ratio
$m_q^2/s$,  where $m_q$ is the mass of the quark and $s$ is
the center of mass energy. 
At LEP, even for the heaviest produced quark, the bottom
quark, we have $(m_b/m_Z)^2 \sim 0.003$ and, 
therefore, for inclusive observables, 
such as total hadronic cross section, or
the decay width of $Z\rightarrow \bar{b}b$, the mass effects
are almost negligible.
The situation is different for more exclusive observables.
It has been stressed~\cite{LO} that quark mass effects 
could be enhanced in jet production cross sections. In this case
observables depend also on a new variable, $y_c$, the jet-resolution 
parameter that defines the multiplicity of the jets.
In fact, these effects have already been seen~\cite{Fuster}. 
However, it has also been shown in~\cite{LO} that
second-order QCD predictions are necessary in order to perform a meaningful
comparison of the theory with experimental results and to possibly extract 
the value of the quark mass from such measurements.

In previous papers~\cite{PL} we have presented the results of the
calculation of some three-jet fractions at LEP
including next-to-leading (NLO) corrections and keeping the full dependence
on the $b$-quark mass.
In this paper we focus on several technical
details that appear in the calculation. 
Our results for some IR divergent integrals
with massive quarks are  new and could
be used in many other calculations of this type.

The main difficulty of the NLO calculation is 
the appearance, in addition to ultraviolet (UV) divergences,
of infrared (IR) singularities because gluons are massless.
The Bloch-Nordsieck (BN) and Kinoshita-Lee-Nauenberg (KLN)
theorems~\cite{BNKLN} assure that jet cross-sections 
are infrared finite and free from collinear divergences. 
The IR singularities of the NLO one-loop Feynman diagrams
cancel against the IR divergences that appear
when the differential cross-section for four-parton production
is integrated over the region of phase-space 
where either one gluon is soft 
or two gluons are collinear. 
Nevertheless, infrared and collinear divergences 
appear in intermediate steps and should be treated 
properly. 
We used dimensional regularization to
regularize both UV and IR divergences~\cite{IR}
because it preserves the QCD Ward identities.

For massive quarks, the gluon-quark collinear singularities, 
that appear in the massless case, are softened 
into logarithms of the quark mass. The number of IR 
singularities is, therefore, smaller. However, quark masses 
complicate extremely the loop and phase-space integrals. 
In particular, diagrams containing 
gluon-gluon collinear divergences are much harder to 
calculate than in the massless case.
Furthermore, infrared singularities appear always at the border
of the integration region.
For virtual  contributions, at the border of the one unit side box
defined by the Feynman parameters.
For contributions coming from the emission of real gluons, 
at the border of the phase-space.
Thus, any simplification is impossible, the full calculation has to
be performed in arbitrary
$D=4-2\epsilon$ dimensions and expansions for
$\epsilon \rightarrow 0$ are allowed only at the end.

Several methods of analytical cancellation of infrared
singularities have been developed in the past.
The most popular are the
{\it phase-space slicing method}~\cite{slice}
and the {\it subtraction method}~\cite{ERT,KN,substra}.
We work in the context of the so-called
phase-space slicing method.
In this case the analytical integration over a thin slice at the
border of the phase-space is performed in $D$-dimensions and
the result is added to the virtual corrections.
The integration over the rest of the phase-space, which is finite,
is done numerically for $D=4$. 

The structure of this paper is as follows.
In section \ref{se:virtual} we consider the most complicated,
IR divergent one-loop box integrals that appear in the calculation of 
the three-jet decay rate of the $Z$-boson into massive quarks at NLO.
We start this section by presenting the calculation of
some simpler three-point scalar integrals, 
which control the IR structure of the box diagrams.
One of the  box integrals involves the three-gluon vertex
and was only known in the case of massless quarks~\cite{D0massless}.
We have  calculated it also for the massive case by using dimensional
regularization. This is one of the main contributions of this paper.


Section \ref{se:real} is devoted to the calculation of
the basic phase-space integrals of the $Z$-decay into four partons 
wich lead to IR (collinear) divergent contributions to the three-jet 
fractions. We used the phase-space slicing method.
When one of the gluons is soft, the calculation is performed 
by imposing a theoretical cut on the energy of the 
gluon, $E_3<w$. 
The $w$-cut should be
small enough to allow for expansions for small $w$ in the analytical 
calculations
and should not interfere with the experimental cuts. On the other hand,
it should be large enough to avoid instabilities in the numerical 
integrations.
For the transition amplitudes containing gluon-gluon collinear
divergences we perform the cut on $y_{34} = 2(p_3\cdot p_4)/s$,
the scalar product of the momenta of the two gluons
normalized to the center of mass energy.
These contributions have to cancel exactly the 
appropriate infrared poles of the virtual contributions.
We show explicitly how these basic integrals share
the same infrared behavior as the one-loop 
scalar functions considered in section \ref{se:virtual}.
Finally, in appendices \ref{ap:hyper} and \ref{ap:ps}, we collect some
properties of the hypergeometric functions and the formulae
that define the three- and the four-body phase-space in 
arbitrary space-time $D$-dimensions.

Throughout all this paper we work with the 
following set of dimensionless Lorentz-invariant variables
\bea
\qquad y_{ij} = 2(p_i\cdot p_j)/s, \qquad r_b= m_b^2/s, \non
\eea
where the $y_{ij}$ are the scalar product of all the possible 
pairs of final state momenta
normalized to $s=q^2$, the total center of mass energy,
$p_1^2=p_2^2=m_b^2$ is the mass of 
the quark/antiquark (in particular the $b$-quark mass)
and particles labeled as 3 and 4 represent the gluons,
$p_3^2=p_4^2=0$.
Energy-momentum conservation imposes the following
restrictions
\bea
y_{12} &=& 1-2r_b-y_{13}-y_{23}, \non \\
y_{12} &=& 1-2r_b-y_{13}-y_{14}-y_{23}-y_{24}-y_{34}.
\eea
in the case of the decays
$Z \rightarrow b\bar{b}g$ and $Z \rightarrow b\bar{b}gg$
respectively.

\section{One-loop integrals contributing to three-parton final states}
\label{se:virtual}

We consider here the IR divergent virtual corrections to the
three-parton decay $q \rightarrow p_1+p_2+p_3$ with all external 
particles on-shell, see fig~\ref{Dloop}.

Using the Passarino-Veltman~\cite{PV79} procedure
we can reduce any set of vectorial or tensorial one-loop 
integrals to n-point scalar functions. Therefore only
scalar integrals~\cite{TV79} have to be computed.
At the NLO the relevant one-loop Feynman diagrams contribute
only through their interference with the lowest order
Born amplitudes. Therefore, we 
will be interested only in the real part of such n-point scalar
one-loop integrals. Although we will not specify it, the following
results refer, in almost all cases, just to the real
part of these functions.

\subsection{Three-point functions}

First we present results for the two three-point functions which 
actually define the non-trivial IR structure of the whole virtual 
radiative corrections for the process. 
As we will show below, the IR poles of the
four-point functions contributing to our process are also controlled
by these simpler three-point functions. Thus, we have to consider the 
following scalar one loop integrals, 
see fig~\ref{Cloop},
\bea
& & \frac{i}{16\pi^2} C03(y_{13}) = \mu^{4-D} \non \\
& & \times \int \frac{d^D k}{(2\pi)^D}  
\frac{1}{k^2 (k-p_3)^2 [(k+p_1)^2-m_b^2]}~,
\eea
and
\bea
& & \frac{i}{16\pi^2} C05(y_{12}) = \mu^{4-D} \non \\
& & \times \int \frac{d^D k}{(2\pi)^D} 
\frac{1}{k^2[(k+p_1)^2-m_b^2][(k-p_2)^2-m_b^2]}~.    
\eea
In general, this kind of functions depend on all the 
masses in the propagators, on the square of the difference
of the two external momenta and on the square of the external 
momenta itself. Since we have fixed the mass and we have
imposed the on-shell condition the only remaining 
relevant arguments of these functions are the two-momenta
invariants $y_{13}$ and $y_{12}$, respectively.

\mafigura{7.5cm}{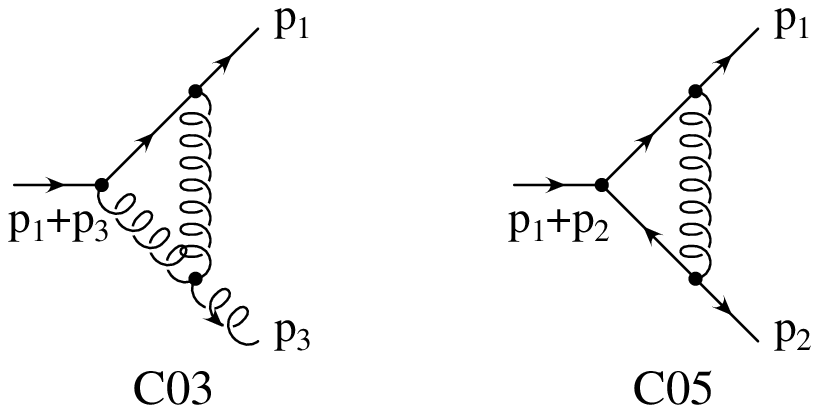}
{Infrared divergent triangle Feynman diagrams.}
{Cloop}

For the $C03$ function, which has a double IR pole, the most 
convenient Feynman parametrization is the one with 
both Feynman parameters running from $0$ to $1$
because then the two integrations decouple. After performing the integration
over the loop momenta we have
\bea
& & C03(y_{13}) = - s^{-1-\epsilon} \mu^{2\epsilon}
(4\pi)^{\epsilon} \Gamma(1+\epsilon) 
\int_0^1 x^{-1-2\epsilon} dx \\
& & \times 
\int_0^1 \left[(1-y)(r_b - y (r_b+ y_{13})) - i \eta  
\right]^{-1-\epsilon} dy~. \non
\eea
We see from here that, as we mentioned, the infrared problem 
is at the border of the Feynman parameter space, in this particular 
case at $x=0$ and $y=1$.
The full calculation has to be performed 
for arbitrary $\epsilon$ values.

We will omit the $i \eta$ term in the following calculations, however,
when needed, we will use the prescription
$r_b \rightarrow r_b - i\eta$, $y_{13} \rightarrow y_{13} + i \eta$
to resolve the ambiguities.

After integration over the x-parameter we are left with 
\bea
& & C03(y_{13}) = s^{-1-\epsilon} \mu^{2\epsilon}
(4\pi)^{\epsilon} \frac{\Gamma(1+\epsilon)}{2\epsilon} 
(r_b+y_{13})^{-1-\epsilon} \non \\
& & \times \int_0^1 (1-y)^{-1-\epsilon} (y_0-y)^{-1-\epsilon} dy~,
\eea
where $y_0=r_b/(r_b+y_{13})$.
The last integral gives rise to a hypergeometric function, \eq{eq:hintegral},
\bea
& & C03(y_{13}) = s^{-1-\epsilon} \mu^{2\epsilon}
\frac{(4\pi)^{\epsilon}}{\Gamma(1-\epsilon)}
\frac{\Gamma(1+\epsilon) \Gamma(-\epsilon)}{2\epsilon} \non \\ 
& & \times (r_b+y_{13})^{-1-\epsilon} y_0^{-1-\epsilon}
{}_2F_1 [1+\epsilon,1;1-\epsilon;\frac{1}{y_0}]~,
\eea
that, after some mathematical manipulations
(see appendix \ref{ap:hyper})
can be written in terms of a dilogarithm function,
\bea
& & {}_2F_1 [1+\epsilon,1;1-\epsilon;\frac{1}{y_0}] = \non \\
& & \left(1-\frac{1}{y_0}\right)^{-1-2\epsilon}
\left[ 1+2\epsilon^2 Li_2\left(\frac{1}{y_0}\right) 
+ O(\epsilon^3) \right]~. \label{eq:hyperprop}
\eea
Above we used \eq{eq:htransform} and \eq{eq:hexp}.
In the region of physical interest $y_0$ is
smaller than one, therefore to extract the real parts we transform
$Li_2(1/y_0)$ in terms of $Li_2(y_0)$ by using the known properties of the
dilogarithm function \cite{DD84}. The ambiguities are solved by using 
the correct $i \eta$ prescription.

The final result we obtain for the real part is,
\bea
C03(y_{13}) &=& s^{-1-\epsilon} \mu^{2\epsilon}
\frac{(4\pi)^{\epsilon}}{\Gamma(1-\epsilon)} 
\frac{1}{y_{13}} \non \\
&\times & \left[ \frac{1}{2\epsilon^2} 
+ \frac{1}{2\epsilon} \log \frac{r_b}{y_{13}^2}
+ \frac{1}{4} \log^2 \frac{r_b}{y_{13}^2} \right. \non \\
&-& \left. \frac{1}{2} \log^2 y_0
- Li_2(y_0) - \frac{7\pi^2}{12} \right]~.
\label{C03}
\eea
Notice that for convenience we have factorized a 
$(4\pi)^{\epsilon}/\Gamma(1-\epsilon)$ 
coefficient because this constant appears also in the four-body
phase-space, see \eq{PSsystem34}.

The $C05$ function has a simple pole and already appeared in the NLO
calculation of two-jet production, although in a different kinematical regime.
We just quote the result since
its calculation is straightforward
\bea
C05(y_{12}) &=& s^{-1-\epsilon} \mu^{2\epsilon}
\frac{(4\pi)^{\epsilon}}{\Gamma(1-\epsilon)}
\frac{1}{(y_{12}+2r_b)\beta_{12}} \non \\ 
&\times& \left[ \left( 
\frac{1}{\epsilon} - \log r_b \right) \log c_{12}
- 2 L(y_{12}) \right]~,
\label{C05}
\eea
where
\bea
L(y_{12}) &=& Li_2(c_{12}) + \frac{\pi^2}{3} 
\non \\
&+& \log (1-c_{12}) \log c_{12} - \frac{1}{4} \log^2 c_{12}~,
\eea
and
\beq
\beta_{12} = \sqrt{1-\frac{4r_b}{y_{12}+2r_b}}~,  \qquad 
c_{12} = \frac{1-\beta_{12}}{1+\beta_{12}}~. 
\eeq

\subsection{Box integrals}

In the one-loop diagrams contributing to the  NLO corrections to the 
three-jet decay rate of the $Z$-boson into massive quarks,
see fig~\ref{Dloop},
we encounter the following set of four-point functions,
\bea
& & \frac{i}{16\pi^2} D05(y_{13},y_{12}) = 
\mu^{4-D} \int \frac{d^D k}{(2\pi)^D} \\ & & 
\frac{1}{k^2 [(k+p_1)^2-m_b^2][(k+p_{13})^2-m_b^2][(k-p_2)^2-m_b^2]}~, \non
\eea
and
\bea
& & \frac{i}{16\pi^2} D06(y_{13},y_{23}) = 
\mu^{4-D} \int \frac{d^D k}{(2\pi)^D} \non \\ & & 
\frac{1}{k^2 (k+p_3)^2 [(k+p_{13})^2-m_b^2][(k-p_2)^2-m_b^2]}~, 
\eea
where $p_{13}=p_1+p_3$.

\mafigura{7.5cm}{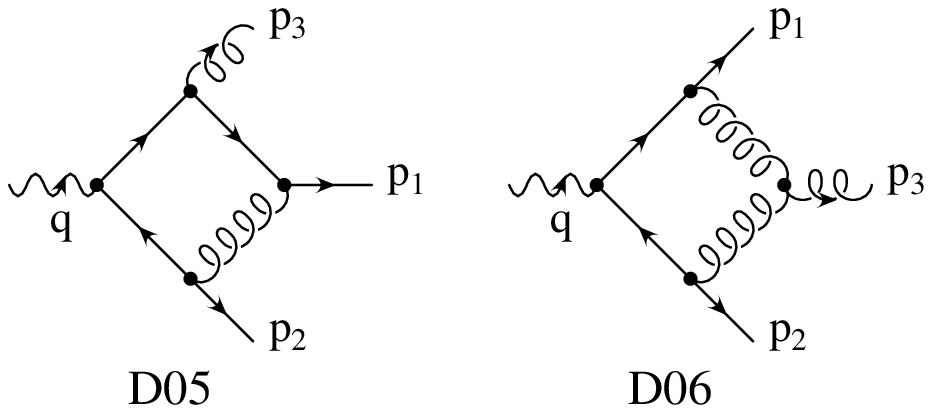}
{NLO box contributions to the three-jet decay rate
of the $Z$-boson into massive quarks.}
{Dloop}

Both four-point functions are IR divergent. The integral $D05$ has
a simple infrared pole, while $D06$ presents double infrared poles
because it involves a three-gluon vertex.
Before the calculation of these integrals we would like to discuss their
IR behavior. 
To do that we just need to split their non on-shell propagator.

For the $D05$ integral we have
\bea
\frac{1}{(k+p_{13})^2-m_b^2} &=& \frac{1}{p_{13}^2-m_b^2} \non \\ 
&-& \frac{k^2+2k \cdot p_{13}}{(p_{13}^2-m_b^2)[(k+p_{13})^2-m_b^2]}~.
\eea
The infrared divergence is isolated in
the first term of the righthand side of the previous equation that
give rise to $C05(y_{12})/y_{13}$, where $C05$ is the three-point
function defined in \eq{C05}.
The other term generates two infrared finite one-loop integrals. 

For the $D06$ function things are quite more complex. 
The splitting of the non on-shell propagator does not give
directly the divergent piece due to the presence of double
infrared poles.
First of all, we have to notice that $D06$ is invariant under the interchange
of particles 1 and 2, i.e., the result of the integral should be 
symmetric in the two-momenta invariants $y_{13}$ and $y_{23}$.
The proper way for extracting the infrared piece
is as follows \cite{kosower}
\bea
D06(y_{13},y_{23}) &=& \frac{1}{y_{13}} C03(y_{23}) \non \\ 
&+& \frac{1}{y_{23}} C03(y_{13}) + \mathrm{finite \, terms}~.
\eea

The box integral, $D05$, comes from an electroweak-like 
Feynman diagram and it has already been calculated~\cite{Denner90}
by using a photon mass infrared regulator.
Since only simple infrared poles 
appear in the $D05$ function, we are allowed to identify
the logarithm of the gluon (photon)  mass, $\log \lambda^2$,
in the result of ref.~\cite{Denner90}, with a pole in 
dimensional regularization:
$\log \lambda^2 \rightarrow (4\pi)^\epsilon/\Gamma(1-\epsilon)/\epsilon$.
Thus, we have
\bea
D05(y_{13},y_{12}) &=& s^{-2-\epsilon} \mu^{2\epsilon} 
\frac{(4\pi)^{\epsilon}}{\Gamma(1-\epsilon)}
\frac{1}{(y_{12}+2r_b)\beta_{12} y_{13}} \non \\ 
&\times& \bigg[ \left( 
\frac{1}{\epsilon} - \log \frac{r_b}{y_{13}^2} \right) \log c_{12} 
\non \\
&+& \pi^2 + \log^2 c - Li_2(1-c_{12}^2) \non \\
&+& 2 Li_2(1-c_{12}~c) + 2 Li_2(1-c_{12}/c)
\bigg]~,
\label{D05}
\eea
with
\beq
\beta = \sqrt{1-4r_b}~, \qquad c = \frac{1-\beta}{1+\beta}~. 
\eeq

To calculate $D06$ we consider the following set of Feynman 
parametrizations
\bea
& & \frac{1}{k^2 (k+p_3)^2} = \int_0^1 \frac{dz}{[k^2+2k\cdot p_3 z]^2}~, 
\non \\
& & \frac{1}{[(k+p_{13})^2-m_b^2][(k-p_2)^2-m_b^2]} = \non \\
& & \qquad \int_0^1 \frac{dy}{[k^2+(2k\cdot q+2p_1\cdot p_3)y
-2k\cdot p_2]^2}~, 
\eea
that again we combine with the help of the identity
\beq
\frac{1}{a^2 b^2} = \int_0^1 dx \frac{6x(1-x)}{[a\, x+b(1-x)]^4}~.
\eeq
After these parametrizations 
our integral becomes linear in the $z$-Feynman parameter
\bea
& & \frac{i}{16\pi^2} D06(y_{13},y_{23}) = 
\mu^{4-D} 
\int_0^1 6 x(1-x) dx\, dy  \non \\ & & 
\times \int \frac{d^D k}{(2\pi)^D}
\int_0^1 \frac{dz}{[k^2-(M_1 z + M_0 (1-z))+i\eta]^4}~,
\eea
where we defined
\bea
M_0 &=& s\, (1-x) 
\left[ r_b \, (1-x) - (1-x) \, y(1-y) - y_{13} xy \right]~,
\non \\ 
M_1 &=& s\, (1-x) 
\left[ r_b \, (1-x) - (1-x) \, y(1-y) - y_{23} x(1-y) \right]~.
\eea
Again we will omit the $i \eta$ term, and,
when needed, we will use the prescription
$r_b \rightarrow r_b - i\eta$,
$y_{13,23} \rightarrow y_{13,23} + i \eta$.


Performing the $z$-integration before the integration over the loop momentum
we obtain
\bea
& & \frac{i}{16\pi^2} D06(y_{13},y_{23}) = 
s^{-1}\mu^{2\epsilon} 
\int_0^1 \frac{2 dx\, dy}{y_{13} y - y_{23}(1-y)} \non \\ & & 
\times \int \frac{d^D k}{(2\pi)^D}
\left\{ \frac{1}{(k^2-M_1)^3} - \frac{1}{(k^2-M_0)^3} \right\}~,
\eea
that results, after integration
over the loop momentum, into 
\bea
& & D06(y_{13},y_{23}) = s^{-2-\epsilon} \mu^{2\epsilon} 
(4\pi)^{\epsilon} 
\Gamma(1+\epsilon) \non \\
& & \times 
\int_0^1 \frac{dy~A^{-1-\epsilon}}{y_{13} y - y_{23}(1-y)}
\int_0^1 d\, x\ (1-x)^{-1-\epsilon} 
\left[1 - x (1+ \frac{y_{13} y}{A})\right]^{-1-\epsilon} 
\non \\
& & \qquad \qquad \qquad \qquad \qquad \qquad \qquad
+ (y_{13} \leftrightarrow y_{23})~,
\eea
where
\beq
 A = r_b-y(1-y)~.
\eeq
After integration over the $x$-Feynman parameter
we get a hypergeometric function
\bea
& & D06(y_{13},y_{23}) = s^{-2-\epsilon} \mu^{2\epsilon} 
(4\pi)^{\epsilon} 
\frac{\Gamma(1+\epsilon)}{-\epsilon} \non \\
& & \times \int_0^1 dy \frac{A^{-1-\epsilon}}
{y_{13} y - y_{23}(1-y)} 
\ {}_2F_1[1+\epsilon,1;1-\epsilon;1+\frac{y_{13}y}{A}] \non \\
& & \qquad \qquad \qquad \qquad \qquad \qquad \qquad 
+ (y_{13} \leftrightarrow y_{23})~.
\eea
Again, using the transformation properties of ${}_2F_1$
and expanding in $\epsilon$, 
we can write the above hypergeometric function in terms of 
the  dilogarithm function, see appendix \ref{ap:hyper}.
Moreover, as
the dilogarithm function contributes at the order $O(\epsilon^2)$
we are allowed to substitute it with its value at $y=0$, $Li_2(1)=\pi^2/6$,
obtaining
\bea
& & D06(y_{13},y_{23}) 
= s^{-2-\epsilon} \mu^{2\epsilon} 
(4\pi)^{\epsilon} 
\frac{\Gamma(1+\epsilon)}{-\epsilon}(1+\epsilon^2\frac{\pi^2}{3})\non \\
& &\times \left( \int_0^1 dy \frac{(-y_{13} y)^{-1-2\epsilon}}
{y_{13} y - y_{23}(1-y)} A^{\epsilon}
+ (y_{13} \leftrightarrow y_{23})~\right).
\eea
Partly expanding in $\epsilon$ to the required order, 
we can rewrite our integral before the last integration in the following form
\bea
& & D06(y_{13},y_{23}) 
= s^{-2-\epsilon} \mu^{2\epsilon} 
\frac{(4\pi r_b)^{\epsilon}}{\Gamma(1-\epsilon)} 
\frac{-1}{\epsilon}(1+\epsilon^2\frac{\pi^2}{2}) 
\non \\
& & \times \int_0^1 dy \left(
\frac{(-y_{13} y)^{-1-2\epsilon}}
{y_{13} y - y_{23}(1-y)} 
-\epsilon \frac{\log(1-y(1-y)/r_b)}
{y_{13} y (y_{13} y - y_{23}(1-y))}\right)\non \\
& & \qquad \qquad \qquad \qquad \qquad \qquad \qquad \qquad \qquad 
+ (y_{13}\leftrightarrow y_{23})~,
\eea
where we have used 
$\Gamma(1+\epsilon)\Gamma(1-\epsilon)=(1+\epsilon^2 \pi^2/6+O(\epsilon^4))$.
The first part of the integrand gives, up to a factor
$-(-y_{13})^{-2\epsilon}/y_{13}/y_{23}/(-2\epsilon)$
a hypergeometric function
that can be reduced to the dilogarithm,
\bea
& & {}_2F_1[1,-2\epsilon;1-2\epsilon;\frac{y_{13}+y_{23}}{y_{23}}] = 
\non \\
& & \qquad
\left(-\frac{y_{13}}{y_{23}}\right)^{2\epsilon}
\left[1+4\epsilon^2 Li_2\left( \frac{y_{13}+y_{23}}{y_{13}}  \right) 
+ O(\epsilon^3) \right]~.
\label{eq:hh}
\eea
Using the above equation and 
the symmetry $y_{13} \leftrightarrow y_{23}$, 
we can rewrite $D06(y_{13},y_{23})$~ as
\bea
& & D06(y_{13},y_{23}) 
= 
s^{-2-\epsilon} \mu^{2\epsilon} 
\frac{(4\pi r_b)^{\epsilon}}{\Gamma(1-\epsilon)}(1+\epsilon^2\frac{\pi^2}{2})  
\frac{1}{y_{13} y_{23}2\epsilon^2}
\non \\
& &
\bigg[(-y_{13})^{-2\epsilon}
      \left(\frac{-y_{23}}{y_{13}}\right)^{-2\epsilon}+
      (-y_{23})^{-2\epsilon}
      \left(\frac{-y_{23}}{y_{13}}\right)^{2\epsilon}+\non\\
& &      
      2\epsilon^2\left(
      2Li_2\left(\frac{y_{13}+y_{23}}{y_{23}}\right)+
      2Li_2\left(\frac{y_{13}+y_{23}}{y_{13}}\right)
     -\int_0^1 dy
\frac{\log(1-y(1-y)/r_b)}{y(1-y)}\right)\left.\right.\bigg]~.
\eea
Performing the last integral,
and fully expanding in $\epsilon$
(in this expansion the $i\eta$ prescription should be used
to avoid ambiguities)
we have the following final result for the real part of the
function $D06(y_{13},y_{23})$
\bea
D06(y_{13},y_{23}) &=& s^{-2-\epsilon} \mu^{2\epsilon} 
\frac{(4\pi)^{\epsilon}}{\Gamma(1-\epsilon)} 
\frac{1}{y_{13}y_{23}} \non \\
&\times & \left[ \frac{1}{\epsilon^2} 
+ \frac{1}{\epsilon} \log \frac{r_b}{y_{13} y_{23}} \right. 
\non \\ &-& \left. \log^2 c 
+ \frac{1}{2} \log \frac{r_b}{y_{13}^2} \log \frac{r_b}{y_{23}^2}
-\frac{3\pi^2}{2} \right]~.
\eea

\section{Infrared divergent phase-space integrals}

\label{se:real}

We consider now the tree-level four-parton decay
$q \rightarrow p_1+p_2+p_3+p_4$.
When at least one gluon is radiated from an external quark (or gluon) line,
the propagator-factors, $1/(p_i\cdot p_j)$, 
can generate infrared and gluon-gluon collinear divergences at the border 
of the four-body phase-space. In this section we integrate these factors in 
a thin slice at the border of the phase-space and we 
show how these basic phase-space integrals are related to the
infrared divergent scalar one-loop integrals of the previous
section.

\subsection{Integrals containing soft gluon divergences}

The phase-space of $n+1$ particles can be written as the product of  
the $n$-body phase-space times the integral over the 
energy and the solid angle of the extra particle.
In arbitrary $D=4-2\epsilon$ dimensions we have
\beq
dPS(n+1) = \frac{1}{2(2\pi)^{D-1}} E^{D-3} \: dE \: d\Omega \: dPS(n)~. 
\eeq
Suppose $E_3$ is the energy of a soft gluon:
$E_3 < w$ where $w$, with $w$ very small, 
is an upper cut on the soft gluon energy. Let's consider
\beq
\int_0^w E_3^{D-3} dE_3 \: d\Omega \frac{1}{(2 p_1 \cdot p_3)^2}~, 
\eeq
where $p_3$ is the momentum of the gluon and $p_1$ is the momentum of
a quark. After integration over the trivial angles we find
($x=\cos \theta$, with $\theta$ being the angle between $\vec{p_1}$ and
$\vec{p_3}$)
\beq
\frac{2\pi^{\frac{D}{2}-1}}{4 \Gamma \left(\frac{D}{2}-1\right)}
\int_0^w dE_3 E_3^{D-5} 
\int_{-1}^1 dx \frac{(1-x^2)^{\frac{D-4}{2}}}{(E_1-{\bf p}_1 x)^2}~. 
\eeq
Here ${\bf p}_1$ stands for the modulus of the threemomentum.
In the eikonal region we can suppose $E_1$ being almost independent
from $E_3$
\beq
E_1 \simeq \frac{\sqrt{s}}{2} (1-y_{24})~, 
\eeq
with $y_{24} = 2 (p_2\cdot p_4)/s$.
The integral over $E_3$ can be easily done and gives
a simple infrared pole in $\epsilon$. Since the $x$ dependent part
is completely finite we can expand in $\epsilon$ before
the integral over this variable is performed.
The final result we get is as follows
\bea
\frac{1}{2(2\pi)^{D-1}} & &
\int_0^w E_3^{D-3} dE_3 d\Omega \frac{1}{(2 p_1 \cdot p_3)^2} = \non \\
& & \frac{1}{16\pi^2} \frac{(4\pi)^{\epsilon}}{\Gamma(1-\epsilon)} 
\frac{1}{E_1^2-{\bf p}_1^2} \non \\
& & \times \left[ - \frac{1}{2\epsilon} + \log 2\omega
+ \frac{E_1}{2{\bf p}_1} 
\log \frac{E_1-{\bf p}_1}{E_1+{\bf p}_1} \right]~.
\label{energy1}
\eea

Let's consider now
\beq
\int_0^w E_3^{D-3} dE_3 d\Omega 
\frac{1}{(2 p_1 \cdot p_3)(2 p_2 \cdot p_3)}~, 
\eeq
with $p_2$ the momentum of the antiquark.
In this case we can combine the denominators as follows
\beq
\frac{1}{(2 p_1 \cdot p_3)(2 p_2 \cdot p_3)} = 
\int_0^1 dy \frac{1}{[2 p_3 \cdot (p_1+(p_2-p_1)y)]^2}~.
\eeq
We get therefore the same integral structure as in \eq{energy1},
but with the fourmomentum 
$p_A = p_1+(p_2-p_1)y$,
instead of $p_1$.
After integration over the y-parameter we get
\bea
\frac{1}{2(2\pi)^{D-1}} & &
\int_0^w E_3^{D-3} dE_3 d\Omega 
\frac{1}{(2 p_1 \cdot p_3)(2 p_2 \cdot p_3)} = \non \\
& & \frac{1}{16\pi^2} \frac{(4\pi)^{\epsilon}}{\Gamma(1-\epsilon)} 
\left[
\left( \frac{1}{2\epsilon} - \log 2\omega \right) \right. \non \\
& & \times \left. \frac{2}{(y_{12}+2r_b) \beta_{12}} \log c_{12}
+ F(y_{14},y_{24}) \right]~. 
\label{energy2}
\eea
Notice that this last integral has the same divergent structure
as the scalar one loop $C05$ function defined in \eq{C05}.
The finite contribution, the $F(y_{14},y_{24})$ function, is rather
involved. We write it in terms of the variables $y_{13}$ and
$y_{23}$ for later convenience
\beq
F(y_{13},y_{23}) = \frac{1}{b \sqrt{1-a^2}}[G(z_1)+G(z_2)-2G(1)]~,
\eeq
where
\bea
& & G(z) = \log t_1 \log \frac{t_1-z}{t_1+z}
+ \log t_2 \log \frac{z+t_2}{z-t_2} \non \\
&+&  \log \frac{t_1-t_2}{2} 
\log \frac{(t_1+z)(z-t_2)}{(t_1-z)(z+t_2)}  \non \\
&+& \frac{1}{2}
\left[ \log^2 (t_1+z) + \log^2 (t_1-z) \right. \non \\
&+& \left. \log^2 (z-t_2) - \log^2 (z+t_2) \right]  \non \\
&-& \frac{1}{4} \left[ \log (t_1+z) + \log (t_1-z) + 
\log \left( \frac{z-t_2}{z+t_2} \right) \right]^2 \non \\
&+&  Li_2 \left( \frac{z-t_2}{z-t_1} \right)
+ Li_2 \left( \frac{z-t_1}{z+t_1} \right) \non \\
&-& Li_2 \left( \frac{z+t_2}{z+t_1} \right)
- Li_2 \left( \frac{z-t_2}{z+t_2} \right)~,
\eea
with 
\bea 
& & a = \sqrt{h_p}/b~, \\
& & b^2 = 1-4r_b-2(1-2r_b)(y_{13}+y_{23}) \non \\
& & \qquad + (1-r_b) (y_{13}^2+y_{23}^2)
+ y_{13}y_{23}(3-2r_b-y_{13}-y_{23})~, \non 
\eea
where $h_p$ is the function that defines the limits of 
the three-body phase-space, see \eq{bornPS}, and
\bea
& & t_{1,2} = (1 \pm \sqrt{1-a^2})/a~,\\
& & z_1 = \exp \left[ \cosh^{-1} 
\left( \sqrt{(1-y_{13})^2-4r_b}/(a (1-y_{13})) \right) \right]~, \non \\
& & z_2 = \exp \left[ \cosh^{-1} 
\left( \sqrt{(1-y_{23})^2-4r_b}/(a (1-y_{23})) \right) \right]~. \non
\eea

\subsection{Integrals containing gluon-gluon collinear divergences}

Let's consider the following phase-space integral
\beq
PS(4) \frac{1}{y_{13}y_{34}}~.
\eeq
We work in the so-called ``3-4 system'', \eq{PSsystem34},
where as usual $p_1$
denotes the momentum of the quark and $p_3$ and $p_4$ are the
two gluon momenta.
This integral has a singularity when either the gluon labeled as ``3'' is
soft or the two gluons are collinear, that is $y_{34}=0$. Then we take
$y_{34}<w$ with $w$ 
very small. We can decompose the four-body phase-space as the 
product of a three-body phase-space in terms of variables 
$y_{134}$ and $y_{234}$ times the integral over $y_{34}$ and 
the two angular variables $v$ and $\theta'$
\bea
PS(4) & & \frac{1}{y_{13}y_{34}} = PS(3)(y_{134},y_{234}) \non \\
& & \times \frac{S}{16 \pi^2}  \frac{(4\pi)^\epsilon}{\Gamma(1-\epsilon)}
\frac{1}{N_{\theta'}} \int_0^{\pi} d\theta' \sin^{-2\epsilon} \theta' \non \\ 
& & \times \int_0^1 dv (v(1-v))^{-\epsilon} \int_0^w dy_{34}
\frac{y_{34}^{-1-\epsilon}}{y_{13}}~,
\eea
where $S=1/2!$ is the statistical factor.
In the ``3-4 system'' the two-momenta invariant $y_{13}$ can be written 
in terms of the integration variables as
\beq
y_{13} = \frac{1}{2} \left( y_{134} - \sqrt{y_{134}^2-4 r_b y_{34}} (1-2v)
\right)~.
\label{y13}
\eeq
The $y_{13}$ factor is independent of the $\theta'$-angle
so we can get rid of this first integral. Moreover, 
$y_{13}$ contains a piece independent of the $v$-variable and 
another one that is odd under the interchange
$v \leftrightarrow(1-v)$. With the help of this symmetry we can 
avoid the use of the square root in \eq{y13} by using the identity
\beq
\frac{1}{2} \left( \frac{1}{g+f(v)} + \frac{1}{g-f(v)} \right)
= \frac{g}{g^2-f(v)^2}~,
\eeq
then, we rewrite our integral as
\bea
& & PS(4) \frac{1}{y_{13}y_{34}} = PS(3)(y_{134},y_{234}) \non \\
& & \times \frac{S}{16 \pi^2}  \frac{(4\pi)^\epsilon}{\Gamma(1-\epsilon)} 
\int_0^1 dv (v(1-v))^{-\epsilon} 
\int_0^w y_{34}^{-1-\epsilon} dy_{34} \non \\
& & \times \frac{2 y_{134}}{y_{134}^2-(y_{134}^2-4 r_b y_{34})(1-2v)^2}~.
\eea
After integration over $y_{34}$ we get the first infrared pole
\bea
 PS(4) \frac{1}{y_{13}y_{34}} &=& PS(3)
\frac{S}{16 \pi^2}  \frac{(4\pi)^\epsilon}{\Gamma(1-\epsilon)}
\frac{-1}{\epsilon} \frac{w^{-\epsilon}}{2 y_{134}} \non \\
& \times & \int_0^1 dv (v(1-v))^{-1-\epsilon}
{}_2F_1[1,-\epsilon;1-\epsilon;-B \frac{(1-2v)^2}{v(1-v)}]~,
\eea
with $B = w (r_b/y_{134}^2)$.
Similarly to \eq{eq:hh} we can write ${}_2F_1$
in terms of a dilogarithm function
\bea
{}_2F_1& &[1,-\epsilon;1-\epsilon;-\frac{a}{b}] = \non \\ 
& & \left(\frac{b}{a+b}\right)^{-\epsilon}
\left[ 1+ \epsilon^2 Li_2\left(\frac{a}{a+b}\right) 
+ O(\epsilon^3) \right]~.
\eea
Thus, we obtain 
\bea
PS(4) \frac{1}{y_{13}y_{34}} &=& PS(3)
\frac{S}{16 \pi^2}  \frac{(4\pi)^\epsilon}{\Gamma(1-\epsilon)}
\frac{-1}{\epsilon} \frac{w^{-\epsilon}}{2 y_{134}}
\int_0^1 dv (v(1-v))^{-1-2\epsilon} \non \\ 
& \times & \left( v(1-v) + B (1-2v)^2 \right)^{\epsilon} \non \\
& \times & \left[ 1 + \epsilon^2 
Li_2 \left( \frac{B (1-2v)^2}{v(1-v)+B (1-2v)^2} \right) +
O(\epsilon^3)\right]~.
\eea
The above integrand is symmetric under the interchange 
$v \leftrightarrow (1-v)$. We perform the integral only 
in half of the integration region and apply 
the following change of variables
\beq
u = 4 v(1-v)~.
\eeq
In addition, since the IR divergence occurs at $u=0$ and the term
proportional to the dilogarithm function is already order $\epsilon^2$,
we can substitute the dilogarithm function by its value at $u=0$, that
is $Li_2(1)=\pi^2/6$. After this, we get
\bea
 PS(4) \frac{1}{y_{13}y_{34}} &=& PS(3)
\frac{S}{16 \pi^2}  \frac{(4\pi)^\epsilon}{\Gamma(1-\epsilon)}
\frac{-1}{\epsilon} 
\frac{w^{-\epsilon} B^{\epsilon}}{y_{134}} 2^{4\epsilon}
\left(1+\epsilon^2\frac{\pi^2}{6}\right) \non \\
&\times & \int_0^1 du\, u^{-1-2\epsilon} (1-u)^{-1/2}
\left[ 1- \left( 1-\frac{1}{4B} \right) u \right]^{\epsilon}~,
\eea
which gives again a hypergeometric function
\bea
PS(4) \frac{1}{y_{13}y_{34}} &=& PS(3)
\frac{S}{16 \pi^2}  \frac{(4\pi)^\epsilon}{\Gamma(1-\epsilon)}
\frac{-1}{\epsilon}
\frac{1}{y_{134}} 
\left( \frac{r_b}{y_{134}^2} \right)^{\epsilon} 2^{4\epsilon}
\left(1+\epsilon^2\frac{\pi^2}{6}\right)
\sqrt{\pi} \non \\
& \times &\frac{\Gamma(-2\epsilon)}{\Gamma(\frac{1}{2}-2\epsilon)}
{}_2F_1[-\epsilon, -2\epsilon;\frac{1}{2}-2\epsilon;
1-\frac{1}{4B}]~.
\eea
The hypergeometric function is already $1+O(\epsilon^2)$
but now it is not possible to write it in terms of a 
dilogarithm function.
If we are interested just in the pole structure we can stop here.
(Observe that neither the double pole or the simple one depend on
the eikonal cut $w$). To get the complete finite part further 
mathematical manipulations must be performed with the
hypergeometric function. We use \cite{AS72}
\bea
{}_2F_1& &[-\epsilon, -2\epsilon;\frac{1}{2}-2\epsilon;
1-\frac{1}{4B}] = \non \\
& & (4B)^{2\epsilon} \Gamma(\frac{1}{2}-2\epsilon) \non \\
& & \times \left\{\frac{\Gamma(\epsilon)}
{\Gamma(-\epsilon)\Gamma(\frac{1}{2})}
{}_2F_1[-2\epsilon, \frac{1}{2}-\epsilon; 1-\epsilon; 4B] \right. \non \\
& & + \left. \frac{\Gamma(-\epsilon)}
{\Gamma(-2\epsilon) \Gamma(\frac{1}{2}-\epsilon)}
{}_2F_1[-\epsilon, \frac{1}{2}; 1+\epsilon; 4B] \right\}~.
\eea
For $w$ small enough, $w \ll y_{134}^2/r_b$, 
we can use the Gauss series \cite{AS72} to expand the
hypergeometric functions around $B \rightarrow 0$.
The final result we obtain is as follows
\bea
& & dPS(4) \frac{1}{y_{13}y_{34}} = dPS(3)(y_{134},y_{234})
\frac{S}{16 \pi^2}  \frac{(4\pi)^\epsilon}{\Gamma(1-\epsilon)}
\frac{1}{y_{134}} \non \\
& & \times \left\{
\frac{1}{2\epsilon^2}-\frac{1}{2} \log^2 w 
+ \left(\frac{1}{2\epsilon}-\log w\right) 
\log \frac{r_b}{y_{134}^2} \right. 
- \frac{1}{4} \log^2 \frac{r_b}{y_{134}^2} 
 - \frac{\pi^2}{4} +O(w)\biggr\}~.
\eea
Notice we get the same infrared poles as in the one-loop 
three point function $C03$, \eq{C03}, if we identify 
$y_{134}$ with $y_{13}$. In the limit $y_{34}\rightarrow 0$
the momentum $p_{34}=p_3+p_4$ behaves as the momentum of 
a pseudo on-shell massless particle since $p_{34}^2\rightarrow 0$
and the function that defines the limits of the four-body 
phase-space, $h_{34}$ in \eq{h34}, reduces to the 
three-body phase-space function $h_p$, see \eq{bornPS}.
Therefore, within this limit, this identification provides 
the key to see the cancellation of the infrared divergences.

\section{Summary}
We have presented the basic ingredients needed to compute the 
second order QCD corrections to the three-jet decay rate of the 
$Z$-boson into massive quarks. This includes, 
for massive quarks, the first calculation of the double infrared
divergent box integral, we called $D06$, in dimensional 
regularization. We have presented results for both the 
infrared part and the finite contributions of some
basic infrared divergent one-loop and phase-space integrals.
We showed they share the same infrared behavior. Therefore,
the cancellation of the infrared divergences is assured
as stated by the BN and KLN theorems \cite{BNKLN}.

\vspace{1cm}

We would like to acknowledge interesting discussions with 
S. Catani, A. Denner, A. Manohar and A. Pich. 
We are also indebted with S. Cabrera, 
J. Fuster and S. Mart\'{\i} for an
enjoyable collaboration. 
M.B. thanks the Univ. de Val\`encia for the warm hospitality during
his visit.
The work of G.R. and A.S. has been supported in
part by CICYT (Spain) under the grant AEN-96-1718 and IVEI.
The work of G.R. has also been supported in part by CSIC-Fundaci\'o Bancaixa.

\appendix

\section{Some properties of the hypergeometric function}

\label{ap:hyper}

Here we present some properties of the hypergeometric function 
\cite{AS72} we use in the main body of the paper.
The integral representation of the hypergeometric function is
\beq
{}_2F_1[a, b;c; z] = 
\frac{\Gamma(c)}{\Gamma(b)\Gamma(c-b)}
\int_0^1 d\, t\  t^{b-1} (1-t)^{c-b-1} (1-t z)^{-a}~.
\label{eq:hintegral}
\eeq
From the above equation one can derive the following transformation
properties,
\bea
{}_2F_1[a, b;c; z] &=&(1-z)^{c-a-b} {}_2F_1[c-a, c-b;c; z] \non\\ 
&=&(1-z)^{-b}{}_2F_1[b, c-a;c; \frac{z}{z-1}]~.
\label{eq:htransform}		   
\eea
Using the Gauss series for the hypergeometric function \cite{AS72}
one can easily see that
\begin{equation}
{}_2F_1[\alpha\epsilon, \beta \epsilon ;1+\gamma\epsilon ; z] =
1+\epsilon^{2} \alpha \beta Li_2(z)+O(\epsilon^3)~,
\label{eq:hexp}
\end{equation}
where, $Li_2(z)$ is the Spence or dilogarithm function
\beq
Li_2(z) = - \int_0^z dt \frac{\log (1-t)}{t}~.
\eeq

\section{phase-space in $D=4-2\epsilon$ dimensions}

\label{ap:ps}

The phase-space for $n$-particles in the final state
in arbitrary $D$ space-time dimensions~\cite{IR}
$(D=4-2\epsilon)$ has the following general form
\bea
dPS(n) &=& (2\pi)^D \prod_{i=1,n} \frac{d^{D-1}p_i}{(2\pi)^{D-1}2E_i}
\delta^D \left( q-\sum_{i=1,n}p_i \right) \non \\
&=&  (2\pi)^D \prod_{i=1,n} \frac{d^{D}p_i}{(2\pi)^{D-1}} \non \\
&\times & \delta(p_i^2-m_i^2)\Theta(E_i) \delta^D 
\left( q-\sum_{i=1,n}p_i \right)~.
\label{phased}
\eea

Let's consider the decay into three particles,
$q \rightarrow p_1 + p_2 + p_3$, where particles 1 and 2 
share the same mass, $p_1^2 = p_2^2 = m_b^2$, and particle 3 is 
massless, $p_3^2 = 0$. 
In terms of the two-momenta invariant variables
$y_{13}=2(p_1 \cdot p_3)/s$ and 
$y_{23}=2(p_2 \cdot p_3)/s$, with $s=q^2$, we get 
\bea
PS(3) &=& \frac{s}{16(2\pi)^3}
\frac{1}{\Gamma(2-2\epsilon)}
{\left( \frac{s}{4\pi} \right) }^{-2\epsilon} \non \\
&\times &\int \theta(h_p) h_p^{-\epsilon} dy_{13} dy_{23}~,
\eea
where the function $h_p$ which gives the phase-space boundary
has the form
\beq
h_p = y_{13} y_{23} (1-y_{13}-y_{23}) - r_b (y_{13}+y_{23})^2~,
\label{bornPS}
\eeq
with $r_b = m_b^2/s$.

For the case of the decay into two massive and two 
massless particles, $q \rightarrow p_1 + p_2 + p_3 + p_4$
with $p_1^2 = p_2^2 = m_b^2$ and $p_3^2 = p_4^2 = 0$,
it is convenient to write the four-body phase-space as a 
quasi three-body decay
\bea
\qquad \qquad q \rightarrow & & p_{34} + p_1 + p_2    \non \\
                            & & \hookrightarrow p_3 + p_4~. \non
\eea
In the c.m. frame of particles 3 and 4 the four-momenta can
be written as 
\bea
p_1 &=& (E_1, \ldots, 0, {\bf p}_1)~, \non \\ 
p_2 &=& (E_2, \ldots, {\bf p}_2 \sin \beta, {\bf p}_2 \cos \beta)~, 
\non \\
p_3 &=& E_3 (1, \ldots, \sin \theta \cos \theta', \cos \theta)~,  \non \\
p_4 &=& E_4 (1, \ldots, - \sin \theta \cos \theta', - \cos \theta)~, \non 
\eea
where the dots in $p_3$ and $p_4$ indicate $D-3$ unspecified,
equal and opposite angles (in $D$ dimensions) and $D-3$ zeros
in $p_1$ and $p_2$. We will refer to this as the
``3-4 system''~\cite{ERT}.

Energies and threemomenta can be written
in terms of the following invariants
\bea
& & y_{34}  = \frac{2 (p_3 \cdot p_4)}{s}~,  \non \\
& & y_{134} = \frac{2 (p_1 \cdot p_{34})}{s}~, \qquad 
y_{234} = \frac{2 (p_2 \cdot p_{34})}{s}~,  \non
\eea
where $p_{34}=p_3+p_4$. We obtain
\bea
E_1 &=& \frac{y_{134}\sqrt{s}}{2 \sqrt{y_{34}}}~, \qquad
{\bf p}_1 = 
\frac{\sqrt{s}}{2 \sqrt{y_{34}}} \sqrt{y_{134}^2-4r_b y_{34}}~, \non \\
E_2 &=& \frac{y_{234}\sqrt{s}}{2 \sqrt{y_{34}}}~, \qquad
{\bf p}_2  = \frac{\sqrt{s}}{2 \sqrt{y_{34}}} \sqrt{y_{234}^2-4r_b y_{34}}~,
\non \\
E_3 &=& E_4 = \frac{\sqrt{y_{34}\, s}}{2}~. \non
\eea
Setting $v=(1-\cos \theta)/2$, the $D$-dimensional 
phase-space in this system is
\bea
& & PS(4) = \frac{s}{16(2\pi)^3}
\frac{1}{\Gamma(2-2\epsilon)}
{\left( \frac{s}{4\pi} \right) }^{-2\epsilon} \int dy_{134} dy_{234}
\non \\
& & \times s^{1-\epsilon}
\frac{S}{16 \pi^2}  \frac{(4\pi)^\epsilon}{\Gamma(1-\epsilon)}
\int dy_{34} \theta(h_{34}) h_{34}^{-\epsilon} y_{34}^{-\epsilon} \non \\
& & \times \int_0^1 dv (v(1-v))^{-\epsilon}
\frac{1}{N_{\theta'}} \int_0^{\pi} d\theta' \sin^{-2\epsilon} \theta' ~,
\label{PSsystem34}
\eea
where $S=1/2!$ is the statistical factor, 
$N_{\theta'}$\index{Ntheta@$N_{\theta'}$} is a
normalization factor
\beq
N_{\theta'} = \int_0^{\pi} d\theta' \sin^{-2\epsilon} \theta' 
= 2^{2\epsilon} \pi 
\frac{\Gamma(1-2\epsilon)}{\Gamma^2(1-\epsilon)}~,
\eeq
and the function
\bea
h_{34} &=& y_{134}y_{234}(1-y_{134}-y_{234})-r_b(y_{134}+y_{234})^2 \non \\
&+& \left( -1+4r_b+2(1-2r_b)(y_{134}+y_{234}) \right. \non \\
&-& \left. y_{134}^2-3y_{134}y_{234}-y_{234}^2 \right) y_{34} \non \\
&+& 2 (1-2r_b-y_{134}-y_{234}) y_{34}^2-y_{34}^3~,
\label{h34}
\eea
defines the limits of the phase-space.
Observe that $h_{34}$ reduces to \eq{bornPS} in the case
$y_{34}\rightarrow 0$. Furthermore, in this limit,
$p_{34}=p_3+p_4$ behaves as the momentum of 
a pseudo on-shell massless particle since $p_{34}^2\rightarrow 0$.

\end{document}